\documentclass[12pt]{iopart}

\usepackage{iopams}  
\usepackage{graphicx}

\begin{document}

\title[Analytic Harmonic Approach to the $N$-body problem]{Analytic Harmonic Approach to the $N$-body problem}

\author{J R Armstrong, N T Zinner, D V Fedorov and A S Jensen}
\ead{jeremy@phys.au.dk}
\address{Department of Physics and Astronomy, Aarhus University, 
DK-8000 Aarhus C, Denmark}

\date{\today}

\begin{abstract}
We consider an analytic way to make the interacting $N$-body problem
tractable by using harmonic oscillators in place of the relevant 
two-body interactions. The two-body terms of the $N$-body Hamiltonian  are approximated by 
considering the energy spectrum and radius of the relevant two-body 
problem which gives frequency, center position, and zero point energy
of the corresponding harmonic oscillator. Adding external harmonic
one-body terms, we proceed to solve the full quantum mechanical $N$-body
problem analytically for arbitrary masses. Energy eigenvalues, eigenmodes, and correlation
functions like density matrices can then be computed analytically.
As a first application of our formalism, we consider the $N$-boson 
problem in two- and three dimensions where we fit the two-body 
interactions to agree with the well-known zero-range model for 
two particles in a harmonic trap. Subsequently, 
condensate fractions, spectra, radii, and
eigenmodes are discussed as function of dimension, boson number $N$,
and scattering length obtained in the zero-range model.
We find that energies, radii, and condensate fraction increase 
with scattering length as well as boson number, while radii 
decrease with increasing boson number. Our formalism is 
completely general and can also be applied to fermions, Bose-Fermi
mixtures, and to more exotic geometries.
\end{abstract}


\maketitle

\section{Introduction}

Determining the ground state and properties of $N$ interacting particles in some fixed geometry is 
at the core of many disciplines in physics and other natural sciences. However, 
in general even for moderate values of $N$, methods based on first principles are either 
intractable or extremely time-consuming. Fortunately, the properties of many 
systems can be described by correlations that involve just a few particles and 
the problem of many particles can be reduced to consider a much smaller number
in the background of the remaining particles. Two-body interactions still 
have a tendency to make even such few-body problems very difficult to solve and 
insights gained from approximations that allow analytic treatments are therefore
very useful.

In configuration interaction methods, where large basis states of properly symmetrized
wave functions are built and diagonalized to determine system properties, a 
basis of harmonic oscillator states can be very convenient as matrix elements of the 
two-body interaction are 
easy to calculate. This approach was used early on in the context of the
nuclear shell model \cite{goep55,heyd90}.
Turning this method upside down the potentially complicated
two-body interaction could be reproduced or simulated by a simple
harmonic oscillator potential. The great advantage is obviously that
the coupled set of differential equations of motion are much simpler
to solve, and a number of properties are easily systematically
obtained as functions of interaction parameters and particle number.

In subfields of physics where the structure is already established
at a given level of accuracy, 
the insight gained from oscillator approximations
is most often insufficient for improvements. 
However, in cold atomic gas physics it is now possible to 
prepare systems with very exotic and often unknown properties, to use
controlled two-body interactions, to study different geometries and
dimensions, and to vary trapping conditions \cite{bloch08}.  
Analytical oscillator approximations can therefore 
be expected to be very valuable, as it
has been in other fields of physics. While we are concerned mainly
with the quantum mechanical many-body problem here, we note that 
similar techniques are currently used also in the study of classical 
dynamics \cite{ludvig2010}.

Obviously, the accuracy of the oscillator approximation increases as
the potentials resemble oscillators. This means that pronounced smooth
potential single minima with room for bound states are directly suited
for investigations of system structures as functions of particle
number and other characteristics.  However, also much weaker
binding potentials would reveal correct overall qualitative, and
perhaps also semi-quantitative, behaviour in an appropriate oscillator
approximation.  This is especially emphasized by the universal
behaviour of a number of weakly bound structures which only depend on
integral properties like scattering length.  
In that case, the bulk part of the potentials are not crucial by themselves but
rather the large distance effect or equivalently the tail of the wave
function or the binding energy.  All these quantities are related in
the correct model descriptions but for specific purposes a subset may
suffice.  It seems clear that continuum properties like scattering
behaviour are beyond the regime where useful results can be expected. Still
even phase shifts can be extracted from properly discretized continuum
states \cite{zhan08}.

The harmonic oscillator has been used in virtually every aspect of
physics where potentials are needed.  Still, in full generality, the
procedure is not well described probably for several reasons. First,
relative and center-of-mass (cm) motions are not separable even for two
interacting particles with different masses in a trapping one-body
potential.  Second, the simplest approximation for a self bound
$N$-body system is found by the mean-field approximation where the
spurious cm-motion is ignored.  Third, one- or two-body potentials
centered at different points in space have only been of interest when
the harmonic approximation is insufficient, as e.g., in chemistry and
for crystal structures.  Now optical lattices and split traps offer the possibility to also 
use multi-centered potentials.  
If only the relative motion is of interest, 
a decoupling scheme is necessary in order to separate the full solution to
relative and cm-motions.

For cold atoms the oscillator approximation has been applied recently
in \cite{brosens97a,brosens97b,brosens97e,brosens97c,brosens97d,tempere98a,lemmens99a,foulon1999,tempere00a}
and in \cite{zalu00,yan03,gajd06}.  These works considered
Bose-condensate properties for equal mass particles, and this
is almost the only case where the center-of-mass motion separates.
In this report we develop formalism to treat 
the most general harmonically interacting
system for bosons (or distinguishable particles). 
In turn, we solve the $N$-body problem for arbitrary
quadratic forms of the one- and two-body interactions. The use of
Cartesian coordinates allows simple solutions for both one, two and
three spatial dimensions, and any non-spherical behaviour of the
corresponding potentials. First we derive the transformation
matrices from initial to final particle coordinates where the
differential equations are completely decoupled.  We also derive
expressions for various quantities like energies, root-mean-square
radii, density matrix and its dominating eigenvalue.  The hope is to
provide simple tools to reveal at least qualitative features of the
new and unknown systems under design and investigations particularly in
the field of cold atomic gases.

We demonstrate our method by finding solutions in two (2D) and three (3D)
dimensions for $N$ identical and pairwise interacting bosons in
external harmonic one-body potentials. The pairwise interactions 
are taken from the celebrated results of Busch {\it et al.} \cite{busch98} 
for two particles with zero-range interaction in a harmonic trap, the
validity of which have been tested in ultracold atomic gas experiments \cite{stoferle06}.
Our method thus provides an analytical approximation to the $N$-boson 
problem with short-range interactions. The condensate fraction is 
readily available from our calculations and shows interesting behaviour
as the scattering length is tuned and the number of particles changes. 
In particular, at small positive scattering length we find a highly 
fragmented state in both two and three dimensions.

\section{Theoretical derivations}

We first define the Hamiltonian in its most general quadratic form
for both coordinates and kinetic energy derivative operators.
The spin degrees of freedom are omitted and symmetries of the spatial
wave functions can in principle easily be imposed by permutations of
the coordinates.  We use matrices to simplify the derivations.
We then derive the coordinate transformation to decouple the set of
coupled oscillators and distinguish between cases where the
cm motion is free and confined by one-body potentials. Lastly, 
we calculate pertinent properties.

\subsection{Hamiltonian}

We consider a system of $N$, possibly different, particles of mass
$m_k (k=1,\dots,N)$  interacting through deformed harmonic potentials $V_{\textrm{int}}$.  The
particles are in addition subject to external one-body potentials,
$V_{\textrm{ext}}$, for each particle constructed as a sum of harmonic
oscillators with different centers.  The total Hamiltonian $H=T+V$
with kinetic energy $T$ and potential $V = V_{\textrm{int}} + V_{\textrm{ext}}$ is then
given by
\begin{eqnarray} \label{e40}
   T &=&  - \sum_{k=1}^{N} \frac{\hbar^2}{2m_k}\Bigg(
 \frac{\partial^2}{\partial x_{k}^2} +
 \frac{\partial^2}{\partial y_{k}^2} +
 \frac{\partial^2}{\partial z_{k}^2}\Bigg) \;\;,  \\ \label{e50}
 V_{\textrm{int}} &=& \frac{1}{4} \sum_{i,k=1}^{N} \Bigg( V_{ik,0 } +  \mu_{ik} 
 \Big(\omega^2_{x,ik}(x_{i}-x_{k} + x_{ik,0})^2 \\ \nonumber
 &+&  \omega^2_{y,ik} (y_{i}-y_{k}+ y_{ik,0})^2 + 
 \omega^2_{z,ik} (z_{i}-z_{k}+ z_{ik,0})^2 \Big) \Bigg) ,
 \\ \label{e60}
 V_{\textrm{ext}} &=&  \frac{1}{2} \sum_{k=1}^{N} m_k \bigg(
 \omega^2_{x,k} (x_{k}-x_{k,0})^2 \\ \nonumber &+& 
 \omega^2_{y,k} (y_{k}-y_{k,0})^2 + \omega^2_{z,k} (z_{k}-z_{k,0})^2 \bigg)\;,
\end{eqnarray}
where $(x_{k},y_{k},z_{k})$ are the $(x,y,z)$-coordinates of the
$k'$th particle, $\mu_{ik} = m_im_k/(m_i+m_k)$ is the reduced mass of
particles $i$ and $k$,
$(\omega_{x,ik},\omega_{y,ik},\omega_{z,ik})$ are the
frequencies in the $(x,y,z)$-directions for the interaction potential
between particles $i$ and $k$, and
($\omega_{x,k},\omega_{y,k},\omega_{z,k}$) are the frequencies in the
$(x,y,z)$-directions for the one-body potential on the $k'$th particle
with centers specified by $(x_{k,0},y_{k,0},z_{k,0})$.  The factor
$1/4$ is made of two factors $1/2$ where one of them is the
conventional notation for an oscillator potential.  The other factor
$1/2$ is to count the two-body interaction only once when the $i,k$
summations are extended to assume all integer values from $1$ to $N$.
The shift of the interaction centers, $x_{ik,0}$, for each pair of the
two-body interactions should then change sign when the particles are
interchanged, $x_{ik,0}= - x_{ki,0}$, which implies that the diagonal
has to vanish, $x_{ii,0}=0$, in accordance with zero self interaction.

This Hamiltonian has the most general quadratic form expressed in
terms of the parameters for one- and two-body oscillator potentials.
The shift of potential energy, $V_{ik,0 }$, of the energy for each
pair allows adjustment without change of structure.  The shifts of
both one- and two-body oscillator centers suggest applications
approximating optical lattice potentials.  The frequencies
traditionally all enter as squares which suggest attraction but the
formalism is equally applicable for imaginary frequencies or
equivalently negative values of these squared frequencies.  To produce
stable solutions with such repulsive interactions requires sufficient
attraction from the other two-body interactions or from the external
fields. The choice of Cartesian coordinates allows independent solution for
each dimension, and thereby treats deformations and different
dimensions without any additional complications. Obviously this also
prohibits direct use of symmetries and conserved quantum numbers where
the dimensions are mixed.  One example is angular momentum
conservation in the absence of external fields.

We now proceed by rewriting the Hamiltonian in matrix form.  For this
we define vectors, $\vec x = (x_{1},x_{2},...,x_{N})^T$, $\vec
\nabla_{x} = (\partial/\partial x_{1},\partial/\partial
x_{2},...,\partial/\partial x_{N})^T$, where a vector is given as a
column of its coordinates, which means the transposed of the row as
indicated by ``$T$''.  The $y$ and $z$-direction can be defined
analogously if necessary.  We treat each coordinate independently and
may therefore omit the $x$-index to simplify the notation in the
derivation.  The $x$-part of the Hamiltonian, $H_{x}$, in
\eref{e40}-\eref{e60} is given by:
\begin{equation}
  H_{x} = \frac{1}{2} \vec{\nabla}_{x}^T T \vec{\nabla}_{x} + 
 \frac{1}{2}\vec{x}^T A\vec{x} + \vec{c}\cdot\vec{x}+V_{\mathrm{shift}} \;,
\label{matrixH}
\end{equation}
where the kinetic energy matrix, $T_{ik}=-\delta_{ik} \hbar^2/(m_i)$,
is diagonal and depends only on inverse masses.  The constant term,
$V_{\mathrm{shift}}$ consists of the sum of all separate shift energies:
\begin{eqnarray}\label{ShiftE}
V_{\mathrm{shift}}&=& \frac{1}{2}\sum_{k=1}^N m_k\omega_{x,k}^2x_{k,0}^2 \\ \nonumber
 &+& \frac{1}{4}\sum_{i,k=1}^{N} \bigg(V_{ik,0} + 
 \mu_{ik}\omega_{x,ik}^2 x_{ik,0}^2 \bigg)  \;\;.
\end{eqnarray}
The quadratic potential term in \eref{matrixH} contains the
symmetric matrix $A$ which is given in terms of masses and frequencies
by
\begin{eqnarray}  \label{e90}
  A_{i \neq k} &=& - \mu_ {ik} \omega^2_{x,ik}  \\
 A_{kk} &=&  m_k \omega^2_{x,k} + \sum_{i, i\neq k}^{N} 
 \mu_{ik} \omega^2_{x,ik}  \;. \label{e100}
\end{eqnarray} 
The components of the coefficient vector $\vec{c}$ in the linear term are
\begin{equation}  \label{e104}
c_k=-m_k\omega_k^2x_{k,0}+\sum_{i=1}^{N}\mu_{ik}\omega_{x,ik}^2x_{ik,0} \; .
\end{equation}

The $y$ and $z$-parts of the Hamiltonian, $H$, are completely analogous
and we have $H =  H_{x} +  H_{y} +  H_{z}$.

\subsection{Reduction to standard form}

The linear term in \eref{matrixH} can be eliminated by translating
the coordinates by 
\begin{eqnarray} \label{e93}
 \vec{x}^{\prime} = \vec{x} - \vec{a} \;\;\;,\;\;\;
\vec{x} = \vec{x}^{\prime} + \vec{a} \;,
\end{eqnarray}
where the translation vector $\vec{a}$ is determined by the
requirement that all terms linear in $x_{k}^{\prime}$ must vanish from
the Hamiltonian.  This condition amounts to $ A\vec{a} = -\vec c$.  As
containing only second derivatives the kinetic energy operator remains
unchanged by this linear translation.  In total the Hamiltonian in the
new coordinates has only the same quadratic terms in both
coordinates and derivatives. However, linear terms have disappeared
and the term, $-1/2\vec{a}^TA\vec{a}$, should be added to $V_{\mathrm{shift}}$
in \eref{ShiftE}.

Any solution, $\vec{a}$, to $A\vec{a} = -\vec c$ eliminates the linear
terms.  Many solutions always can be found, but a unique solution only
exists when $A$ is non-singular, that is $A$ is invertible.  A
singular $A$-matrix is equivalent to a subset of non-interacting
particles which also all are unaffected by the one-body
potentials. Then the corresponding degrees of freedom are already
decoupled and following the trivial motion in free space.  As already
decoupled they should therefore from the beginning be removed from the
reduction procedure.  The dimension of the $A$-matrix is
correspondingly reduced.

In practice, only one frequently occurring example is interesting and
requiring special attention.  This is when all interactions are
translation invariant and only depending on coordinate differences between
particles.  Then the center-of-mass motion is as in free space.
External one-body fields acting on all particles act on the
center-of-mass coordinate and remove the corresponding degeneracy.

Previous works with oscillators \cite{zalu00,yan03,gajd06} considered
equal mass systems where separation of the center-of-mass motion is
straightforward. Here we provide a general procedure valid for all
sets of masses and all one- and two-body interactions. In all cases we
transform to relative and center-of-mass, $X$, coordinates, where the
latter is defined by
\begin{equation}  \label{e114}
 X M = \sum_{k=1}^{N} m_k x_k  \;\;\;,\;\;\;   M = \sum_{k=1}^{N} m_k \;,
\end{equation}
where $M$ is the total mass of the system.  Choosing the new set of
coordinates, $\vec{\tilde{x}}$, by $\tilde{x}_{i} \equiv x_i-X$, for
$i=1,2,...,N-1$, supplemented by $\tilde{x}_{N} \equiv X$, we define the
transformation matrix, $F$, by
\begin{equation}  \label{e119}
\vec{x}^T = F \vec{\tilde{x}}^T,
\end{equation}
where the transformation matrix takes the form
\begin{eqnarray}
F=\left( \begin{array}{cccc}
1 & 0 & \ldots &1 \\
0 & 1 & \ldots &1 \\
\vdots &\vdots &\ddots &1 \\
-\frac{m_1}{m_N}& -\frac{m_2}{m_N}& \ldots& 1 
\end{array} \right). \label{e117}
\end{eqnarray}
The specific labelling singles out the $N'$th (last) particle with
mass, $m_N$, in the transformation $F$.  The inverse transformation,
$F^{-1}$, from original to new coordinates is explicitly given by
\begin{eqnarray}
F^{-1}=\left( \begin{array}{cccc}
1-\frac{m_1}{M} & -\frac{m_2}{M} & \ldots &-\frac{m_N}{M} \\
-\frac{m_1}{M} & 1-\frac{m_2}{M} & \ldots & -\frac{m_N}{M}\\
\vdots &\vdots &\ddots &-\frac{m_N}{M} \\
+\frac{m_1}{M}& +\frac{m_2}{M}& \ldots& +\frac{m_N}{M} 
\end{array} \right). \label{e123}
\end{eqnarray}

The Hamiltonian is now easily transformed to the new set of
coordinates, by direct insertion of \eref{e117} for the
coordinates and the inverse and transposed, $(F^{-1})^T$, for the
derivative kinetic energy operators, $\vec{\nabla}_{\tilde{x}}$.  The
kinetic energy immediately separates into a sum of relative and
center-of-mass dependent terms, whereas the potential part separates
when it is translationally invariant, and otherwise not.  In any case
these coordinates can be used in the following derivations.

\subsection{Decoupling the oscillators}

We assume that we have the quadratic form without linear terms.  We
transform to relative and center-of-mass coordinates in all cases even
though we could have worked with the initial coordinates.  We rename
the coordinates and omit the primes in the following expressions.  We begin by
diagonalizing the quadratic part of the potential, which then is the
$F^TAF$ matrix (see \eref{matrixH}, \eref{e90} and \eref{e100}).  The
orthonormal coordinate transformation $Q$ corresponding to the
diagonalization is defined by the requirements
\begin{eqnarray}  \label{110}
 Q^{T} F^T A FQ = D \;\;, \;\; \vec x = F \vec{\tilde{x}} = FQ \vec t \;,
\end{eqnarray}
where  $D$ is diagonal with eigenvalues $d_i$. 

If $A$ is singular at least one of the eigenvalues, $d_i$, is zero, as
when the interactions only depend on relative coordinates. One of
these zero eigenvalues then has an eigenvector corresponding to the
center-of-mass coordinate.  This mode now has to be fully decoupled
after the kinetic energy is expressed in the same coordinates. This is
achieved in general by replacing the zero eigenvalue by any finite
number.  If only relative motion is of interest the Hilbert space
spanned by this eigenvector should simply be removed. 

We proceed to perform a new non-orthonormal transformation of the
coordinates defined by $\vec t = \bar D \vec u$, where $\bar D$ is
diagonal and given by $\bar D_{ik} = \delta_{ik} \sqrt{d_0/d_i}$.  The
number $d_0$ can be chosen arbitrarily.  We choose to maintain the
total norm which implies that $\Pi_{i=1}^{N} d_i^{(x)} = d_0^{N}$.
This transformation is nothing but scaling the lengths of each of the
eigenvectors to let $\bar D$ become proportional to the unit matrix.  The
transformation from $x$ to $u$-space of the derivative vector is then
given as $\vec{\nabla}_x = (F^{-1})^T Q (\bar D^{-1})^T \vec{\nabla}_u$.
The kinetic energy matrix in the $u$-coordinates is $\bar D^{-1} Q^T
(F^{-1}) T (F^{-1})^T Q (\bar D^{-1})^T$, where the space
corresponding to the center-of-mass coordinate decouples from all the
other degrees of freedom.  This kinetic energy matrix is diagonalized,
i.e. 
\begin{eqnarray} 
\nonumber   T_x  &=& \frac{1}{2}
 \vec{\nabla}_u^T \bar D^{-1} Q^T F^{-1} T (F^{-1})^T Q (\bar D^{-1})^T 
 \vec{\nabla}_u \nonumber  \\ &=& \frac{1}{2} \vec{\nabla}_v^T P^{T} 
 \bar D^{-1}  Q^T F^{-1} T (F^{-1})^T Q (\bar D^{-1})^T  P \vec{\nabla}_v 
 \nonumber \\ &=&  \label{e130} 
\frac{1}{2}  \vec{\nabla}_v^T \bar T \vec{\nabla}_v  \;,
\end{eqnarray}
where the orthonormal transformation $\vec{\nabla}_u = P
\vec{\nabla}_v$, or $\vec u = P^T \vec v$, is chosen to make $\bar T$
diagonal, i.e.  $\bar T_{kk} = - \hbar^2/(\bar m_k)$, or equivalently
diagonalize $\bar D^{-1} Q^T F^{-1} T (F^{-1})^T Q (\bar D^{-1})^T$.

This final orthonormal transformation leaves the potential energy part as
a diagonal matrix with unchanged diagonal elements, i.e.
\begin{eqnarray} 
 V_x &=& \frac{1}{2} \vec x^T F^T A F\vec x  =
  \frac{1}{2} \vec t^T D \vec t = 
 \frac{1}{2} \vec u^T   \bar D^{T} D \bar D\vec u  \nonumber  \\ &=& 
  \frac{1}{2} d_0 \vec u^T  \vec u = \frac{1}{2} d_0 \vec v^T  \vec v 
  = \frac{1}{2}  \vec v^T \bar V_x \vec v  \; ,  \label{e120} 
\end{eqnarray}
where $\bar V_x$ is the diagonal unit matrix with elements $d_0=\bar
m_k \bar{\omega}_k^2$, which defines the output frequencies, $\bar{\omega}_k$.  This is a factorization return to the ordinary
oscillator notation by using the masses determined from the kinetic
energy eigenvalues in \eref{e130}.  If in the process of transforming from the original $\vec x$- cordinates to the $\vec v$ coordinatees zero eigenvalues were generated, then care must be taken to make sure sure that they do not contribute to any final observable.  This is easily
achieved with $\bar{\omega}_k = 0$ for these modes.

The total Hamiltonian for the $x$-coordinate is now a set of
decoupled oscillators in the new coordinates $\vec v$, i.e.
\begin{eqnarray}  \label{e140} 
 H_x &=& V_{\mathrm{shift}}   \\ \nonumber 
 &+&  \sum_{k=1}^{N} \Big(- \frac{\hbar^2}{2\bar m_{x,k}} 
 \frac{\partial^2}{\partial  v_{x,k}^2} + \frac{1}{2} 
 \bar m_{x,k} \bar{\omega}_{x,k}^2 v_{x,k}^2\Big) \;,
\end{eqnarray}
where we inserted the label $x$ on masses, oscillator parameter and
$v$-coordinates.  The harmonic oscillator eigenvalues and
eigenfunctions correspond to the frequencies $\bar{\omega}_{x,k}$ and
the length parameter $b_{x,k}$ as usual given by $b_{x,k}^2 = \hbar/(\bar
  m_{x,k} \bar{\omega}_{x,k})$.  

In total the transformation $M$  from initial to new coordinates and vice
versa are
\begin{eqnarray}  \label{e143}
 \vec v &=& M (\vec x - \vec a) \;,\;\;\;  \vec{x}-\vec a=  M^{-1}\vec v  \;,
  \\  \label{e143a}  M &=& \bar D^{-1} Q^{T}F^{-1}  \;.
\end{eqnarray}
The normal mode of the system is expressed by the eigenvector, and the
corresponding eigenvalue indicates the ease or difficulty of exciting
that particular normal mode.

\subsection{Basic properties}
Observables expressed as expectation values of operators $O$ are found
by
\begin{eqnarray} \label{e158}
 \langle  \Psi | O |\Psi\rangle &=& \int 
 d^N \vec x d^N \vec y d^N \vec z \\ \nonumber && 
  \Psi^*(\vec v_x,\vec v_y,\vec v_z) O(\vec x,\vec y,\vec z)
 \Psi(\vec v_x,\vec v_y,\vec v_z) \; ,
\end{eqnarray}
where the wave functions are simplest in the transformed coordinates
while the operators probably are simpler in the original particle
coordinates. 
We have only considered the spatial wave function $\Psi$. Effects of
spin dependence of interaction or quantum statistics have to be
inserted separately.

The wave functions are products of the three one-dimensional harmonic
oscillator wave functions in the new coordinates, that is Gaussians,
$\exp(-v_{x,k}^2/(2b_{x,k}^2))$, times Hermite polynomials with
arguments $v_{x,k}/b_{x,k}$ for each contributing mode $k$. Other
analogous products arise from the $y$ and $z$-directions.  The
normalization is as usual when we use the final ($v_{x,k}$) coordinates
in the wave function.  All the above transformations, including that
of the non-orthonormal matrix, $\bar D$, was chosen as total norm
conserving. Therefore the volume elements have the same structure in
initial and new coordinates, i.e.  $\Pi_{k=1}^{N} dx_k = \Pi_{k=1}^{N}
d v_{x,k}$.

The expectation value of the Hamiltonian gives the energy, which for
the eigenmode, $k$, of given quantum number, $n_{x,k}$, is $\hbar
\bar{\omega}_{x,k} (n_{x,k} + 1/2)$. Adding the similarly obtained
results from the $y$ and $z$-direction we get the total energy for a
given set, $(n_{x,k},n_{y,k},n_{z,k})$, of quantum numbers
\begin{eqnarray}  \label{160} 
 E_{n_{x,k},n_{y,k},n_{z,k}} &=& 
 V_\mathrm{shift} + \hbar \sum_{k=1}^{N} \big(\bar \omega_{x,k}
 (n_{x,k} + 1/2) \\ \nonumber &+& \bar \omega_{y,k} (n_{y,k} + 1/2) +
 \bar \omega_{z,k} (n_{z,k} + 1/2)\big) \;,
\end{eqnarray}
where the frequencies of the non-contributing modes are inserted as
zero. 

Beside energies the simplest observables are sizes, i.e. 
average values of the interparticle distance(s) in the system. The spatial
extension of the probability for each particle is measured by its root
mean square radius.  With the relation in \eref{e143} we find for the
ground state $\Psi$
\begin{eqnarray} \label{e187}
 && \langle  \Psi | x_i^2 |\Psi\rangle = 
   \langle  \Psi |  \bigg((A^{-1} \vec c)_i 
 + \sum_{k=1}^{N }( F Q \bar D P^{T})_{ik} v_k  \bigg)^2 |\Psi\rangle 
 \nonumber \\ && = 
((A^{-1} \vec c)_i)^2 +
   \sum_{k=1}^{N } \frac{1}{2}b_ {x,k}^2 ((F Q \bar D P^{T})_{ik})^2 \; ,
\end{eqnarray}
where we used that the odd powers of the coordinates vanish after
integration. If the center of mass coordinate $X$ is decoupled we
exclude that term in the summation, and the resulting sum is the
computation of the expectation of $(x_i-X)^2$.  The other dimensions
should be added.

\subsection{One-body density matrix}

The simplest correlation function is the one-body density matrix,
$\rho(x_1,x_1')$, which is the starting point of the calculation of
statistical properties. It is interesting in itself but serves here
also as a simple illustration of analytical calculations. The one-body
density matrix is relevant for bosons (or distinguishable particles)
which is our main application of the method to be discussed in the 
subsequent section. For fermions, the one-body density matrix 
is a rather uninteresting object due to the Pauli principle and 
off-diagonal long-range order (as arising for instance from pairing) 
is determined by the two-body density matrix \cite{yang60}.  
While we do not consider fermions explicitly in the current
presentation, we address the general issue
of symmetrization in the subsection below for completeness.
Note here that
we have taken great care to remove the center-of-mass coordinate so
that only intrinsic coherence is studied. This is particularly important
if $N$ is very small \cite{zinner08} or if the system is subjected to
a temporally or spatially varying additional potential \cite{pit00}.

The ground state wave function is a product of Gaussians in the new
coordinates, $\vec v$ with different units of length $b_{k}$. The
exponent can be written $-\vec v^T B \vec v$ where the matrix $B$ is
diagonal with elements $B_{kk}=1/(2b_{k}^2)$.  Using \eref{e143}
we return to the initial coordinate where 
\begin{equation}
\ln \Psi = - \vec v^T B \vec v = \vec x^T Z \vec x + \textrm{constant} \;,
\end{equation}
which defines the matrix $Z$ when we assume all interaction centers
are at the origin.  The exponent of the density matrix for identical
bosons, where we select the particle labeled $1$, is then
\begin{eqnarray}\label{densstart}
  &-& x_1Z_{11}x_1 - x_1'Z_{11}x_1'  \\
 &-& (x_1+x_1')\sum_{k=2}^N(Z_{1k}+Z_{k1})x_k+2\sum_{i,k=2}^Nx_iZ_{ik}x_k.\nonumber
\end{eqnarray}
To integrate over all other coordinates (from $-\infty$ to $+\infty$)
than $x_1$ and $x_1'$ we complete the squares.  We define the vector
$\vec{w}=Z_{1k}+Z_{k1}$, for $2 \leq k \leq N$ and make the
substitution $\vec x = \vec f - \vec q$, where $\vec q =
1/2(x_1+x_1')(\bar{Z}+\bar{Z}^T)^{-1}\vec w$, and $\bar Z$ is the $Z$ matrix
without the first row and column.  After integration we are left with
the density matrix
\begin{eqnarray}
\rho(x_1,x_1')&=&\mathcal{N}\exp\left[-(x_1Z_{11}x_1+x_1'Z_{11}x_1')\right.\\
&&+\frac{1}{4}(x_1+x_1')^2\vec{w}^T(\bar{Z}+\bar{Z}^T)^{-1} \vec w] \;, \nonumber 
\end{eqnarray} 
where the normalization, $\mathcal{N}$ is
\begin{equation}
\mathcal{N}=\sqrt{\pi(Z_{11}-\frac{1}{2}\vec{w}^T (\bar{Z}+\bar{Z}^T)^{-1}\vec{w})}.
\end{equation}
Going further, we can re-write the exponent as  
\begin{eqnarray} \label{rhorewrite}
  &-& \frac{1}{2} Z_{11} (x_1-x_1')^2  \\ &+&   \nonumber 
\bigg(\frac{1}{4} \vec{w}^T (\bar{Z}+\bar{Z}^T)^{-1} \vec w 
- \frac{1}{2} Z_{11}\bigg) (x_1+x_1')^2 \;,
\end{eqnarray}
where the ratio $d_x$ is given as
\begin{eqnarray} \label{e163}
 d_x = \frac{2 Z_{11}}{ 2 Z_{11}-\vec{w}^T (\bar{Z}+\bar{Z}^T)^{-1} \vec{w} } \; .
\end{eqnarray}
This ratio determines the largest eigenvalue, $\lambda$, obtained after
diagonalization of the density matrix, i.e.
\begin{equation}
\lambda=\frac{2}{1+\sqrt{d_x}},
\end{equation}
where the subscript $x$ is to remind us that this expression is valid
for one dimension.  In more dimensions the wave functions are products
and consequently also the density matrix and its eigenvalues. Thus for
example in three dimensions, we have
\begin{equation}
\lambda=\left(\frac{2}{1+\sqrt{d_x}}\right)\left(\frac{2}{1+\sqrt{d_y}}\right)\left(\frac{2}{1+\sqrt{d_z}}\right).
\label{lambdaeq}
\end{equation}
The size of this eigenvalue is the established measure for the content
of condensate in the wave function.  The remaining eigenvalues can be found in terms of the largest one:
\begin{equation}
\lambda_n=\lambda\left(1-\lambda\right)^n,
\label{denseigval}
\end{equation}
where $n$ is a non-negative integer \cite{gajd00}.  We thus see that if $\lambda$ is close to
one then all other eigenvalues will fall off very fast, whereas for small $\lambda$ one has a distribution of many non-zero eigenvalues and a highly fragmented state.

\subsection{Symmetries imposed by boson and fermion statistics}

The solutions discussed above 
are all obtained without any requirements of symmetry
due to groups of identical particles in the system.  The hamiltonian
necessarily must commute with correponding permutation operators, and
the solutions should then have the proper symmetry. It is then only a
question of selecting those states among the complete set of all
solutions.  Unfortunately, this tempting conclusion is wrong in
general because it is based on the assumption of non-degenerate
states.  Degeneracies allow mixing of different symmetries, which
implies that the solutions are linear combinations of the required
symmetries.  Thus, a simple method to restore the required symmetry is
to construct linear combinations of the available solutions with the
same energy.

We have assumed that the hamiltonian is independent of intrinsic
properties like spin of the particles. This means that the total
wavefunction factorizes into intrinsic (including particle spins) and
spatial parts. Here we only consider the symmetries of the spatial
part which subsequently has to be combined with the remaining parts of
the wavefunction.  The task is not easily formulated in general since
the system may consist of different groups of particles each with
their own symmetry requirement.  This could be a combination of
identical bosons and fermions, or identical particles placed in
different geometries by external fields and consequently effectively
behaving as non-identical particles.

To illustrate the method we assume that all particles are identical,
or alternatively we omit reference to the non-identical particles left
in their orbits while only considering those explicitely mentioned in
the following formulation.  We write the symmetrized or
antisymmetrized spatial wavefunction $\Phi$
\begin{equation} \label{e124}
\Phi(\vec r_{1},\vec r_{2},...,\vec r_{N})  = N_p  \sum_{p} \pi_{p}
 \Psi(\vec r_{p(1)},\vec r_{p(2)},...,\vec r_{p(N)}) \; ,
\end{equation}
where $\vec r_{i} =(x_i,y_i,z_i)$ is the coordinate for particle $i$,
$p$ is a permutation of the set of numbers $\{1,2,...,N\}$, $N_p$ is a
normalization constant and $\pi_{p}$ is $+1$ for spatial (boson)
symmetry, and $\pi_{p} = \pm 1$ for even and odd permutations $p$,
respectively, when we construct wavefunctions with spatial (fermion)
antisymmetry.  The energy of $\Phi$ is given as the energy,
$E_{n_{x,k},n_{y,k},n_{z,k}}$ from Eq.(\ref{160}), which is the same
for each term of $\Phi$.  This conclusion is due to the fact that the
hamiltonian is invariant under these coordinate changes, and the
energy of the state in Eq.(\ref{e124}) is the expectation value of the
hamiltonian.  The wavefunction in Eq.(\ref{e124}) may be identically
vanishing, e.g. when a symmetric, $\Psi$, is antisymmetrized or vice
versa, when an antisymmetric, $\Psi$, is symmetrized. In these cases
the state, $\Phi$, does not describe any physical system.

The procedure can be specified in more details with matrix
manipulations.  If the coordinate transformation corresponding to one
of the permutations in Eq.(\ref{e124}) for one of the Cartesian
coordinates, $x$, is described by $P_x$, i.e. $\vec x_p = P_x \vec x$ 
(with $\vec x = (x_{1},x_{2},...,x_{N})^T$), we
can express the equivalent transformation in the final coordinates by
$\vec v_p = M P_x M^{-1} \vec v$, with $M$ defined in
Eq.(\ref{e143a}).  Thus by replacing $\vec v$ by $\vec v_p$ in the
arguments of the terms in the wavefunctions in Eq.(\ref{e124}) the
same variables $\vec v$ are used as in the unpermuted term.  The same
coordinates are then used to express all terms in the full
(anti)symmetrized wavefunction

The permutations leave the exponent in the wavefunction unchanged,
i.e.  $\vec v_p^T B_p \vec v_p = \vec v^T B \vec v$.  The reason is
that the $N-1$ oscillator frequencies, masses and lengths are
identical. The last length is related to the center of mass motion,
which either has to be omitted without confining field or remains
invariant under the permutations as decoupled from the intrinsic
motion described by the $N-1$ frequencies.  The Hermite polynomials
change arguments but remain polynomials of the same order in the new
coordinates $\vec v_p$. In total, each term in Eq.(\ref{e124}) is a
different polynomial of the same order in the coordinates
corresponding to the permutation.  The expectation value of an
arbitrary operator is then very difficult to find due to the many
terms. Operator expectations are then analytic if they can be found in
an oscillator basis, and in particular this is the case for any
polynomial structure of the coordinate or momentum operators.  
However, in general the expectation values may consist of many terms.

Small excitations leave only few different oscillator quanta in the
wavefunction. The non-vanishing terms in Eq.(\ref{e124}) are then
limited to rather few. The extreme case is the state with all
$n_{x,k}=0$ which is completely symmetric under all permutations, that
is only one term describing the ground state for identical bosons. This
case will be our main concern in the remaining part of the paper.
When all quanta are equal to zero except one with an arbitrary
$n_{x,k} \neq 0$, the number of different permutations are $N$, which
is a manageable number.

Another practical limit is when the number of identical particles
effectively is small, e.g. when identical particles are confined by
external fields to different states or spatial locations.  Then only
very few permutations are present, and the brute force method is
easily applicable. In any case the brute force methods are only
necessary when information beyond the energy is required.  More
suitable procedures can no doubt be designed for specific systems and
corresponding operators, as for instance discussed recently
in the context of the hyperspherical harmonic approach \cite{gattobigio2009}.
However, all derivations can still be made
analytically if the operator expectations are calculable for
oscillator states.

\section{Bosons interacting in a trap} 

Two particles interacting in a trap is obviously the simplest
non-trivial system of $N$ interacting particles.  Still very
interesting features for two particles were discovered for the extreme
limit of a contact interaction and a confining trap potential
\cite{busch98}.
It would not be surprising if an oscillator approximation has difficulties in a
quantitative description of such systems. On the other hand this is a
challenging problem, and the structure of
$N$ particles in a trap with such pairwise interactions are an 
active field of research.  We shall
therefore attempt to extract systematic overall features within our
analytical formalism.  The
philosophy is to reproduce the two-body properties as close as
possible, and then systematically calculate properties of the
many-body system.

Several other approaches have been used for few-body systems 
of atoms interacting via the contact interaction. 
Among these are effective field theory 
approaches \cite{stet07,alhassid08,rotu10,stet10}, shell-model \cite{zinner09,cem09}, 
and Monte Carlo calculations \cite{carlson03,chang07,gezerlis08} in three 
dimensions for fermions, various hyperspherical 
and variational approaches in three \cite{ole02,ole03,ole04a,han06,thoger07} 
and in two dimensions for bosons \cite{lim80,ole04b,blume05,chri09}, and 
exact diagonalization in two dimensions for fermions \cite{ront09,liu10} and
for bosons \cite{liu10}.
The zero-range interaction 
does not allow the diagonalization of the many-body Hamiltonian 
in spatial dimensions greater than one, 
so all these methods must address that in some way.
Most often, this manifests as a cut-off to a particular subspace, 
which effectively renormalizes the interaction strength \cite{stet07,ront08,zinner09}.  
In \cite{chri09}, a Gaussian form of the interaction is used to 
approximate the zero-range potential, and only repulsive interactions 
are considered. With the exception of \cite{ront09}, all discuss energy 
results as a function of model space size, and do not yet discuss other observables.

\subsection{Adjusting the oscillator parameters}

First we must decide how to choose the parameters in the oscillator
model. For two identical bosons we have initially five parameters,
i.e. mass, interaction frequency, energy shift, confining
frequency, and center position of the confining field.  One of these can
always be chosen as a scale parameter or a unit without consequences
for any of the properties.  In the present case, the external field is
provided by its frequency, $\omega_0$, independent of the two-body interaction,
which then leave the interaction frequency, $\omega_{ik}=\omega$ for all $i,k$, and the
energy shift, $V_{ik}=V$ for all $i,k$, to be determined. These parameters all identical 
since we consider indistinguishable particles. We can also immediately 
set the center position of the confining field to zero, as there is no such 
center shifting or multi-center effect in the external field in the original two-body problem.

Our strategy will be to reproduce the ground state properties of the 
model of Busch {\it et al.} \cite{busch98} as much as possible using
a harmonic oscillator. We work exclusively in the
domain where the lowest state is the molecular branch that 
represents a bound state when the 
trap is removed. The population of this particular state was considered 
previously in \cite{bert06,bert07}.  
In three dimensions, this means that we consider only the positive scattering length
side of the resonance,
whereas in two dimensions this branch is always present. This procedure
provides our effective mapping from the two-body interaction to the
solvable $N$-body problem. We note that our choice of branch means
that for large scattering lengths the two-body energy goes 
to $1/2\hbar\omega_0$ (three dimensions) and $\hbar\omega_0$ (two dimensions), where $\omega_0$ is the external field frequency.  For small scattering lengths it goes as the inverse of the scattering length squared in both cases, since
it represent the universal bound state that is also present when $\omega_0\rightarrow 0$.

The Hamiltonian, solved in \cite{busch98} for two particles, is 
\begin{eqnarray} \label{H3D}
H =-\frac{\hbar^2}{2m}\nabla_r^2 +\frac{1}{2}m\omega_0^2r^2 
 +  \frac{4\pi\hbar^2a}{m}\delta^{(3)}
 (\mathbf{r})\frac{\partial}{\partial r}r \; ,
\end{eqnarray}
where $m$ is the mass of the particles, $\omega_0$ is the external
trap frequency as mentioned above, $r$ is the relative coordinate $\mathbf{r}=\sqrt{1/2}(\mathbf{r}_1-\mathbf{r}_2)$ and $a$ is the scattering length of the two-body potential assumed to be a regularized $\delta$-function.

The solutions are given as eigenvalue equations and corresponding wave
functions, $\psi$.  The form of $\psi$ for both two and three
dimensions is found to be
\begin{eqnarray} \label{e227}
\psi(\mathbf{r})\propto\frac{1}{2}\pi^{-D/2}e^{-r^2/(2\ell^2)}
 \Gamma(-\nu)U(-\nu,D/2,r^2/\ell^2) \; ,
\end{eqnarray}
where the dimension is $D=2,3$, the length $\ell$ is given by
$\ell^2=2\hbar/(m\omega_0)$, and the relative energy
$E_{rel}/(\hbar\omega_0)=2\nu+D/2$ is given in terms of the
non-integer quantum number, $\nu$.  The eigenvalue equations are
respectively
\begin{eqnarray}
2\frac{\Gamma(-\nu)}{\Gamma(-\nu-1/2)} &=& \frac{\ell}{a}, \quad D=3
\label{BuschE3D}  \\ \label{e213}
\gamma+\frac{1}{2}\psi(-\nu) &=& \ln\left(\frac{\ell}{a}\right),\quad D=2 \;,
\end{eqnarray}
where $\gamma$ is the Euler-Mascheroni constant.  For a given
scattering length and trap frequency $\nu$ is obtained and both
energies and wave functions are determined.  There are many solutions to the above equations, but we repeat that we work exclusively with the lowest molecular bound state, corresponding to the lowest solution for $\nu$.

Pertinent features from these solutions are now used to choose the
oscillator parameters.  First we directly choose the same external
frequency $\omega_0$ for both particles.  Second, we compute the mean
square radii for $D=2,3$ from \eref{e227} and equate to the
corresponding oscillators, i.e.
\begin{equation}\label{freq}
 \frac{\langle\psi|r^2|\psi\rangle}{\langle\psi|\psi\rangle} 
 = \frac{D\hbar}{2\mu\sqrt{\omega_{}^2+\omega_0^2}}  \;.
\end{equation}
This determines the oscillator frequency $\omega_{}$. Finally, we
adjust the energy shift for the oscillator model to reproduce the
correct two-body energies, i.e.
\begin{equation} \label{e243}
(2\nu+D/2)\hbar\omega_0=\frac{D}{2}\hbar\sqrt{\omega_{}^2+\omega_0^2}+ V_{} \;,
\end{equation}
where $\nu$ is obtained by solving the relevant eigenvalue equation from
\eref{BuschE3D} and \eref{e213}.  The energy shift is
$V_{\textrm{shift}}(N=2)= V_{}$ as seen in
\eref{ShiftE}.

The interaction frequency combined with the trap frequency determines
all structure.  The size of the system is crucial and we have chosen
to reproduce the radius. It is not as meaningful to adjust to energies
since the oscillator cannot be expected to describe very weakly bound
and spatially extended structures.  Still we expect to get an
indication of the energy variation with $N$ from the shift in the
energy zero point.  All oscillator parameters are now determined for
identical bosons, and we can proceed to investigate consequences for
the many-body system.

We note the method to determine the oscillator parameters for the two-body 
potential can be considered much more general and other states may be used, 
like for instance a higher excited state in the zero-range model. In that case 
certain technical considerations arise, for example with respect to the nodal
surfaces of the many-body wavefunction in relation to the two-body interactions
from which it is built. This is particularly important in the case of fermions 
due to the requirement of antisymmetry. A discussion of these questions will be
presented elsewhere.

\subsection{The $N$-body system}

The total energy shift for $N$ particles is simply the number of pairs
times the two-body shift, i.e.
\begin{equation}
V_{\textrm{shift}}=\frac{1}{2}N(N-1)  V_{} \; . 
\label{EnergyShift}
\end{equation}
The external frequency is $\omega_0$ for all particles.  Solving the
oscillator model leads to a set of frequencies where one of them
equals the external trap frequency and corresponds to the decoupled
center-of-mass motion. The remaining $N-1$ other frequencies are
degenerate and for each direction given by
\begin{equation}
\bar{\omega}_{x}= \sqrt{N/2}\sqrt{\omega_{}^2+2\omega_0^2/N} \; .
\label{outputfreq}
\end{equation}
The total ground state energy of the relative motion is then
\begin{equation}
 E_{gs}  = \frac{1}{2}N(N-1)  V_{} + \frac{D}{2} \hbar (N-1) 
 \bar{\omega}_{x}  \; .
\label{e303}
\end{equation}
For repulsive interactions the sign of $\omega_{}^2$ in
\eref{outputfreq} and \eref{e303} is changed. If
$\bar{\omega}_{x}$ becomes imaginary the system becomes unstable,
i.e. if $N \omega_{}^2 < - 2 \omega_0^2$, then the pairwise
repulsion is too strong for the external field to confine the system.

The energy expression in \eref{e303} for a gas of many identical
bosons has $N$-dependent terms of different origin.  The term
proportional to $N^2$ is solely from the interaction, whereas the
$N^{3/2}$ term originates from kinetic energy, two-body interaction,
and external one-body potential, as given by the solution.  The
relative influence of the external potential decreases with $N$.  The
remaining part still varies as $N^{3/2}$ which is a compromise between
the pairwise interaction increasing as $N^2$ and the linear kinetic
energy depending on $N$.

We can compare with energy relations directly derived for a gas of
bosons that interact via a $\delta$-function two-body potential \cite{lovelace87,baym96}, 
which in
the limit of weak interaction in three dimensions is
\begin{equation} \label{e353}
E\simeq \frac{3N}{2}\hbar\omega_0+\frac{N^2U_0}{2(2\pi)^{3/2}b^3} \;\;,\;\;
U_0=\frac{4\pi\hbar^2a}{m} \;,
\end{equation}
where $b$ is the external trap length.  This expression has the same
$N^2$ scaling from the interaction as $V_{\textrm{shift}}$ in the oscillator
model.  The proportionality factor in $V_{\textrm{shift}}$ is obtained through
the frequency and strongly depends on which state is used for the
adjustment. A more detailed comparison is then less direct.  The
linear term in \eref{e353} is obviously arising from kinetic
energy and external field which for the oscillator leads to the term
proportional to $N^{3/2}$.

For a strongly interacting system, where $Na/b\gg 1$, a variational
calculation where part of the kinetic energy is neglected gives
a different dependence \cite{lovelace87,baym96}
\begin{equation}
E=\frac{5}{4}\left(\frac{2}{\pi}\right)^{1/5}
 \left(\frac{Na}{b}\right)^{2/5}N\hbar\omega_0 \;.
\end{equation}
The overall energy scales as $N^{7/5}$ which presumably should be
compared to our result when $\omega_{} \ll \omega_0$ where we get a
linear energy scaling from kinetic energy and part of the two-body
interaction. Another part of the interaction is still scaling as
$N^2$.  A similar result is not surprisingly found with Thomas-Fermi
approximation, where the kinetic energy term is ignored, but where
higher order terms can be included to improve the description \cite{fu03,zinner09b,thoger09}.

The mean square distance from the center-of-mass is also calculated to
be
\begin{equation} \label{e313}
 \langle\left(\mathbf{r}-\mathbf{R}\right)^2\rangle=
 \frac{D(N-1)}{2N^{3/2}}\frac{\hbar}{m}
  \frac{1}{\sqrt{\omega_{}^2/2+\omega_0^2/N}}\;,
\end{equation}
where $\mathbf{R}$ is the location of the center-of-mass.  This
measure of the spatial extension is a reflection of the inverse
behaviour of energies and mean square radii.

\subsection{Energies and radii}
 
\begin{figure}
\centering
\includegraphics[width=0.7\textwidth]{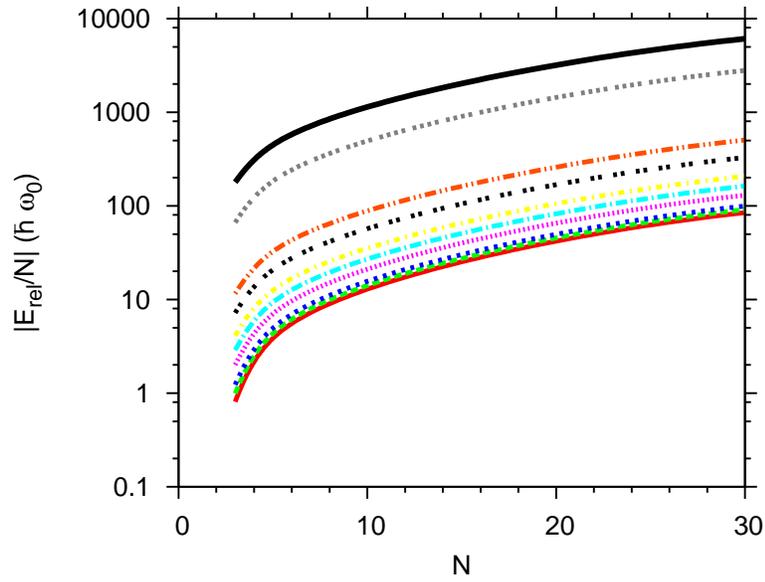}
\caption{Relative energies per particle at several different values of the 3D scattering length as a function of particle number.  The different values of the ratio are, from bottom to top: $a/\ell=100,10,5,2,4/3,1,2/3,1/2,1/5,$ and $1/10$. }
\label{3DErel}
\end{figure} 

The dependence on particle number $N$ is very explicit, but still two
or rather three terms compete with their different $N$-scaling.  We
first show the relative energy per particle (i.e., the energy of the relative motion of the particles without the center-of-mass contribution
corresponding to one particle moving in the external field) in \fref{3DErel} as function of $N$ for three dimensions.  It happens that a two-body system is ``unbound'' in the
oscillator model corresponding to positive relative energy.  This
occurs when the positive contribution in \eref{e243} is larger
than the negative $V_{\textrm{shift}}$.  However, $V_{\textrm{shift}}$ must dominate as
$N$ increases, and in fact we find that once another particle is
added, the relative energy is less than zero for any of the studied
scattering lengths.  The $N$-dependence is very smooth and steadily
increases the binding which varies strongly when the scattering
length approaches zero in units of the trap length.  This is also the region where the magnitude of the two-body bound state energy increases rapidly.

\begin{figure}
\centering
\includegraphics[width=0.7\textwidth]{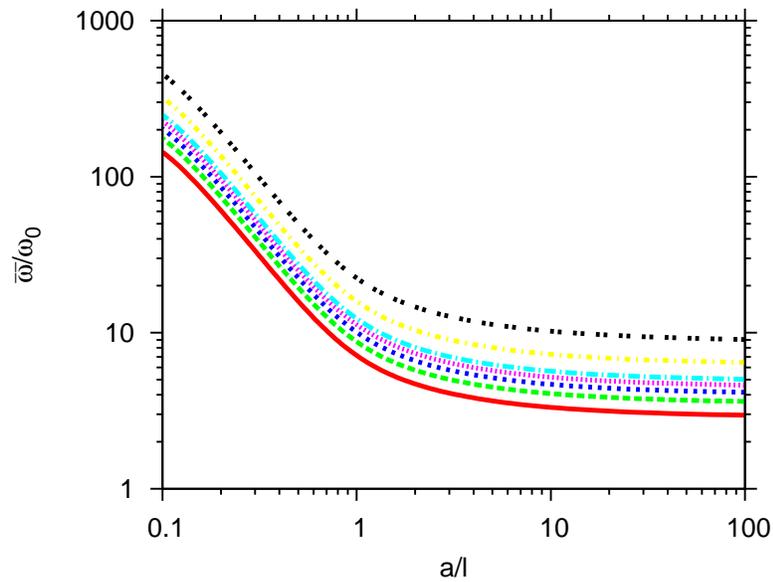}
\caption{Degenerate frequency, $\bar \omega$ of the $N$-particle systems in 3D plotted logarithmically as a function of the ratio of the scattering length to the external confinement length.  The different values of $N$ plotted are, from bottom to top: 3, 4, 5, 6, 10, 20, and 30.}
\label{freq3d}
\end{figure} 

The frequency dependent term in the energy proportional to $\hbar
\bar{\omega}_{x}$ is simply the zero point motion of an oscillator.
It is therefore also the smallest unit of excitation of the system
corresponding to one particle lifted in one dimension from the ground
to first excited state. This is the energy of the normal mode of
excitation.  The dependence is shown in \fref{freq3d} as
function of scattering length from the Busch {\it et al.} model in 
\eref{H3D}. The frequency is
small and constant for scattering lengths larger than the trap length,
while it begins to grow very quickly as soon as the trap length
exceeds the scattering length.

\begin{figure}
\centering
\includegraphics[width=0.7\textwidth]{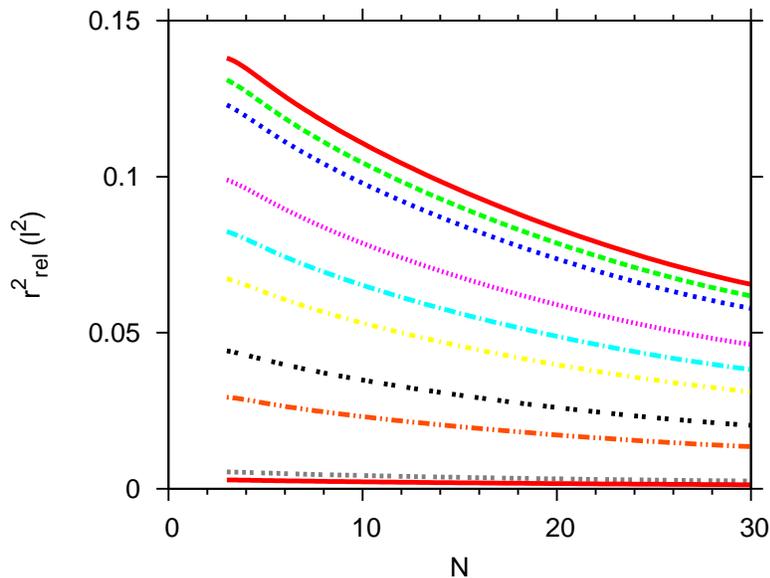}
\caption{Relative sizes, $(r-R_{cm})^2$, at several different values of the 3D scattering length as a function of particle number.  From top to bottom, the ratios are $a/\ell=100,10,5,2,4/3,1,2/3,1/2,1/5,$ and $1/10$.}
\label{3Dsize}
\end{figure} 

The size of the system is, along with the energy, the most fundamental
property.  The intuitive implication that smaller radii follow larger
binding is also observed in \fref{3Dsize}, regardless of the
size of the scattering length, though at the weaker interactions the
change is rather flat for the first few added particles.  The
$N$-dependence is again rather simple as seen in \eref{e313}
where the mean square radius decrease with $1/\sqrt{N}$ for large
$N$. Otherwise the radii are varying with frequency as usual. The
interesting part is rather that the sizes increase substantially with
scattering length for given particle number.

\begin{figure}
\centering
\includegraphics[width=0.7\textwidth]{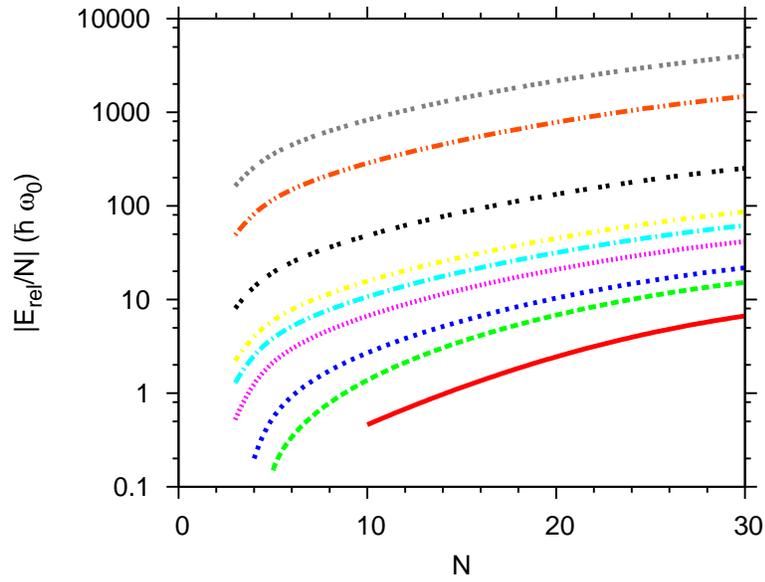}
\caption{Relative energies per particle at several different values of the 2D scattering length as a function of particle number.  From bottom to top, the ratios are $a/\ell=100,10,5,2,4/3,1,1/2,1/5,$ and $1/10$.  Note that a few of the bottom lines at large scattering length over trap length ratios appear to stop at small particle numbers.  This is because there is a sign change in the energy, which is discussed around \eref{e373}.}
\label{2DErel}
\end{figure} 

We now repeat the procedure in two dimensions. The major difference is
obviously from the adjusted oscillator input parameters. The energy
and radius expressions are already given in \eref{e303} and
\eref{e313}. The $N$-dependence of the energy is shown in
\fref{2DErel} where the same overall structure as in
\fref{3DErel} appear.  As in 3D, the energy decreases monotonically with the addition of more atoms for all scattering lengths.

\begin{figure}
\centering
\includegraphics[width=0.7\textwidth]{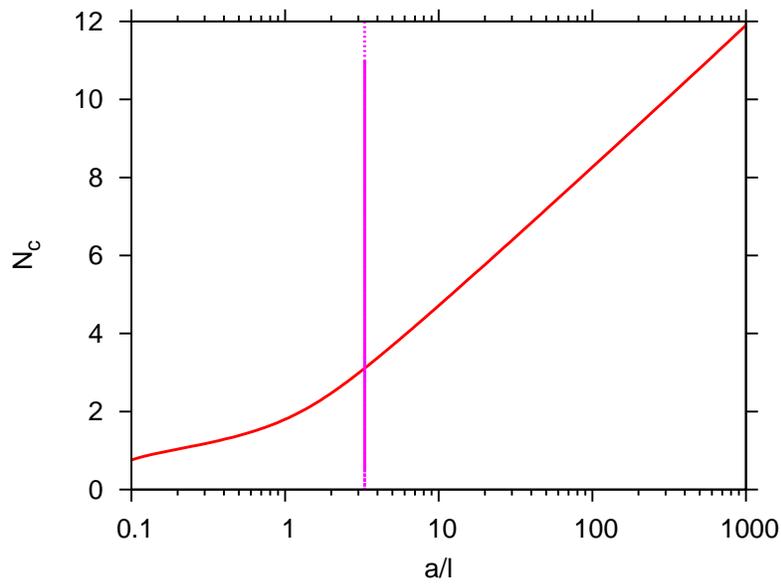}
\caption{The critical number $N_c$ for 2D as function of the ratio of
  the scattering length to the external length.  The vertical line is $a/\ell=3.3$, at which point (to the left of this line) all systems of more than two particles become self-bound.}
\label{Ncrit}
\end{figure} 

The two-body energy shift is less negative because the size
requirement to determine the frequency better matches the one-body
frequency.  Then several of the $N$-body energies are positive,
although ``binding'' (negative energy) finally occurs by adding a
sufficient number of particles.  This critical number, $N_c$, where
the system becomes self bound is given by
\begin{equation} \label{e373}
 N_c = \frac{(\hbar\omega)^2}{V_{}^2}\left[1 + \sqrt{1  
+  4 \frac{\omega_{0}^2}{\omega^2}\left(\frac{V}{\hbar\omega}\right)^2}\right]\; ,
\end{equation}
which for the two dimensional case depends strongly on the initial
scattering length as seen in \fref{Ncrit}.  For the largest
scattering length in \fref{Ncrit} $N_c$ is about $12$, whereas
there is binding for all particles for scattering lengths smaller than about three.  As the scattering length becomes arbitrarily large, the critical number also increases, meaning that the addition of any number of particles is not sufficient to bind the system at infinite scattering length.

\begin{figure}
\centering
\includegraphics[width=0.7\textwidth]{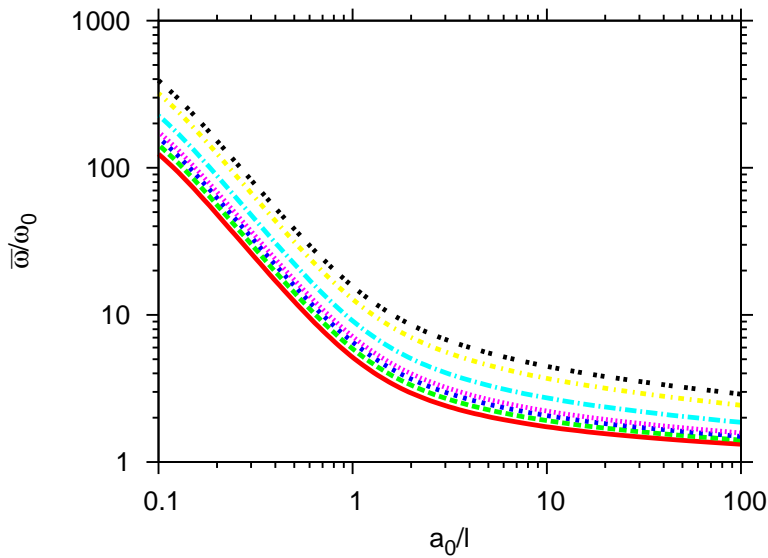}
\caption{Degenerate frequencies, $\bar{\omega}$, of the $N$-particle systems in 2D plotted logarithmically as a function of the ratio of the scattering length to the external confinement length.  The particle numbers plotted are, from bottom to top, 3, 4, 5, 6, 10, 20, and 30.}
\label{freq2d}
\end{figure}

In \fref{freq2d} we show how the degenerate frequency for ten
particle behaves as a function of the scattering length.  The behaviour
is very similar to then 3D result, with both curves turning up strong
after the scattering length becomes less than the external confinement
length.

\begin{figure}
\centering
\includegraphics[width=0.7\textwidth]{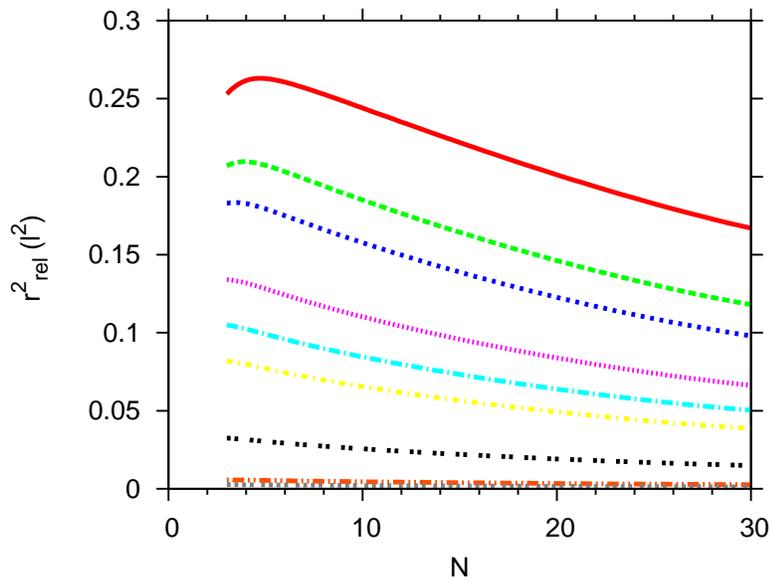}
\caption{Relative sizes at several different values of the 2D scattering length as a function of particle number.  From top to bottom, the ratios plotted are $a/\ell=100, 10, 5, 2, 4/3, 1, 1/2, 1/5, 1/10.$}
\label{2Dsize}
\end{figure}

\Fref{2Dsize} shows the size results in two dimensions.  These
also follow a slightly different trend than in three dimensions.  In
two dimensions, for several scattering lengths, the size actually
increases for the addition of a particle (going from three to four
particles), before falling for all additional particles.  The
scattering lengths where this behaviour is seen correspond roughly to
those that have positive three-body energies.  For smaller scattering
lengths, the relative sizes decreases monotonically with an increase
in particle number.

\subsection{One-body density}

\begin{figure}
\centering
\includegraphics[width=0.7\textwidth]{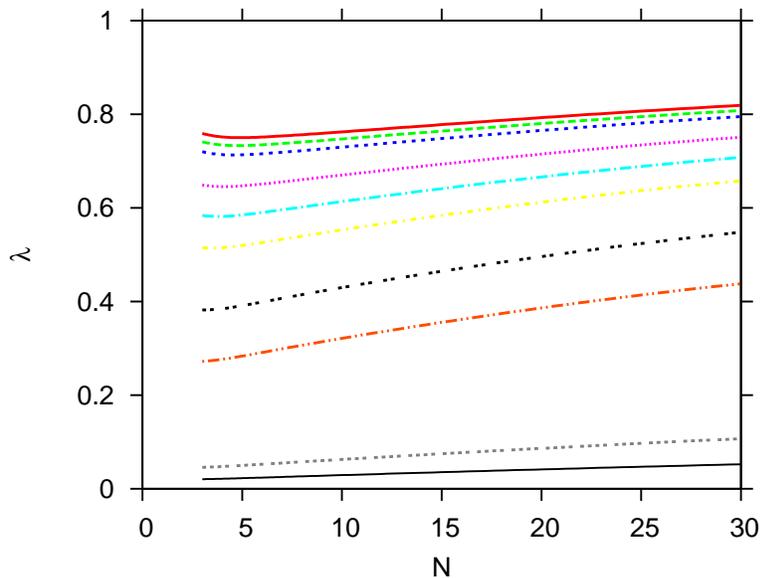}
\caption{The value of $\lambda$ at several different values of the 3D scattering length as a function of particle number.  From top to bottom, the ratios plotted are: $a/\ell=100, 10, 5, 2, 4/3, 1, 2/3, 1/2, 1/5, 1/10.$}
\label{3Dlambda}
\end{figure}

The one-body density matrix has information about the mean-field
content of the wave function \cite{yang60}. This is quantified by the eigenvalue in
\eref{lambdaeq}, which directly measures how much this state has the
structure of a coherent (condensed) state.  In \fref{3Dlambda}, we see how this
eigenvalue, $\lambda$, evolves for 3D with interaction strength and
particle number.  For the most part it increases with particle number,
though at weaker interactions there is a small minimum around five or
six particles and it increases thereafter, while it decreases
uniformly with interaction strength.

The overall increase with $N$ is in agreement with the theorem that
the mean-field wave function is approached as $N$ tends to infinity \cite{yang60}.
The condensate fraction is large for large scattering lengths where
the external field is decisive, and consequently favors the
corresponding mean-field structure. This follows from the fact noted above
that the two-body wave functions become essentially non-interacting oscillator
states given by the confinement ($\omega$ becomes small).
On the other hand, for small
scattering lengths the structure is far from that of a condensate. The
particles are much more tightly bound with strong correlations. Again this
is consistent with the fact that we have very strongly bound two-body 
states in this limit that are very different from the non-interacting harmonic
states.

\begin{figure}
\centering
\includegraphics[width=0.7\textwidth]{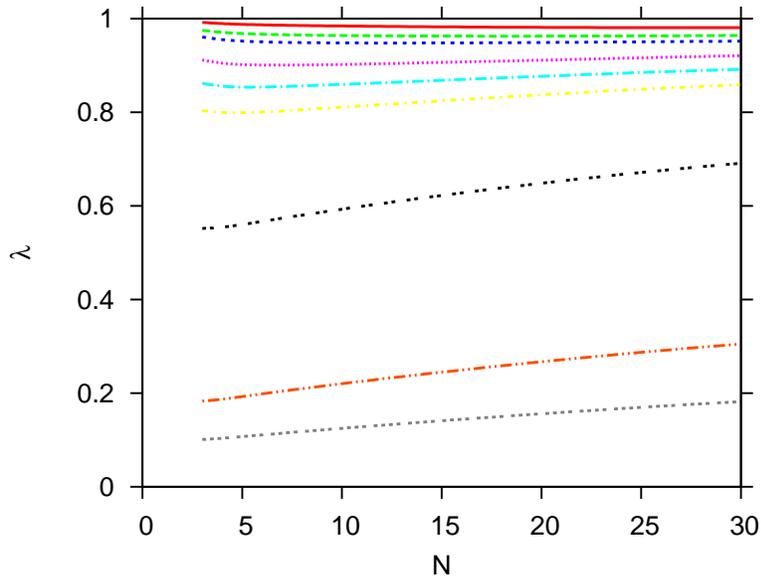}
\caption{Condensate fraction at several different values of the 2D scattering length as a function of particle number.  From top to bottom, the ratios plotted are: $a/\ell=100, 10, 5, 2, 4/3, 1, 1/2, 1/5, 1/10.$}
\label{2Dlambda}
\end{figure}

The question of whether a true condensate actually
exists in a realistic quasi-2D setup with harmonic trapping potentials 
below a certain critical temperature \cite{bagnato91,bloch08} will not be addressed
here. We simply take the appearance of a large eigenvalue in the one-body density
matrix as our working definition of a condensate as in the 3D case above.
The condensate fraction are shown in \fref{2Dlambda} for the 2D
system.
The same tendency of increase with $N$ is present here.  Again
the condensate fraction is large (small) for large (small) scattering
lengths compared to the size of the external field. This reflects the
amount of correlation in the wave function precisely as for 3D.
Quantitatively we find that $\lambda$ is even flatter as a function of
particle number than in 3D, and it is also consistently higher for the
same scattering length. We note that the 2D results consistently produce
larger condensate fractions than in 3D. As we discuss below this 
is connected with the fact that the 2D case resembles a non-interacting
system when $a\rightarrow\infty$ better than the 3D case does. The remaining
deviation of $\lambda$ from unity is due to the separation of center-of-mass.

\subsection{Normal modes and symmetries} 

The characteristic properties of a system are reflected in the
structures and energies of the normal modes. Here we discuss the energies
of the degenerate frequencies which
is the amount of energy required to excited the modes. Unfortunately, the
degeneracy is itself an obstacle for understanding the uncoupled
structures of single excitations. The reason is that any
wave function expressed as a linear combination of the same energies is
equally well qualified. The set of normal modes only has to be
orthogonal, but that can be achieved in infinitely many ways.  One
general exception is if other conserved quantum numbers have to be
restored by specific linear combinations of the degenerate states.
Angular momentum is a prominent example in the absence of external
fields.

We now discuss what determines the degeneracy, or
equivalently what would break it.  First, if all masses, all two-body
interaction frequencies, and all one-body frequencies are equal, then
$N-1$ degenerate frequencies are produced along with the one-body
frequency, which is returned unchanged.
If masses are changed, then the degeneracies are broken, and if $N-2$
masses are changed then all degeneracies will be broken.  This is the
same for the one-body frequencies (those in \eref{e60}), if $N-2$
of them are different from each other, then all the output frequencies
will be different.

For the interaction frequencies in \eref{e50} the situation is
slightly more complex.  If one row and column in the $A$-matrix
of \eref{e90} have the same interaction frequency, i.e., if this
particle interacts the same with all other particles, then there will
be at least one pair of degenerate frequencies.  In general, of the
$N(N-1)/2$ interaction frequencies (off-diagonal elements of the $A$-matrix), 
if $(N-1)(N-4)/2$ ($N>4$) of them are
different, then that is enough to guarantee all symmetries are
destroyed. 
However, all degeneracies can be broken with a wiser
distribution of the different frequencies.  If, for example, the
frequencies immediately above the main diagonal of $A$ are all different and all
the remaining ones being the same (a total of $N$ different off-diagonal
frequencies), then that is enough to destroy all degeneracies of the
resulting normal modes.

Thus, degeneracy can be broken or reached through many different
paths. The structure of the resulting degenerate normal modes depends
on the chosen path.  We still find it interesting to investigate
specially selected normal modes.  If all particles are distinguishable
it should in principle be possible to observe corresponding
vibrational structures where each particle is detected.  To probe the
underlying structure, revealed by the normal modes, we therefore
approach the degenerate limit from an entirely non-degenerate system.
The two simplest ways are to differentiate the particles by minute differences in mass, or alternatively in one-body frequency.

\begin{figure}
\centering
\includegraphics[width=0.7\textwidth]{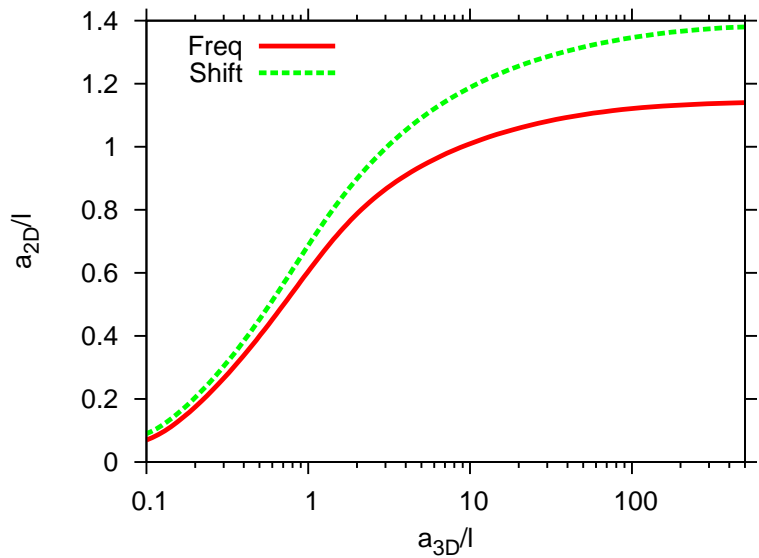}
\caption{The relation between the 2D scattering length to external 
length ratio to the same quantity in 3D for both the interaction 
frequency (red curve) and energy shift (green curve).}
\label{scatlength}
\end{figure}

We first notice that the normal modes are the results of transforming
the set of oscillators to diagonal form. The energies and eigenvalues
are obtained from the matrix depending on masses, external field
frequency, and two-body interaction frequency. Thus by choosing all
these parameters to be identical the normal modes in two and three
dimensions are the same. This is achieved by the same masses, same
external field, and corresponding choices of scattering lengths in 2D
and 3D such that the two-body interaction frequencies from \eref{outputfreq} are identical.  The correspondence is seen in
\fref{scatlength} where both scattering lengths are small at the same time
and also increase simultaneously.  The $2D$ scattering length
approach an upper limit of about $1.14 \ell_{2D}$ for increasing
$a_{3D}$. The size in $2D$ is much smaller than in $3D$.
In \fref{scatlength} we also show the relation between 2D and 3D
oscillator shifts obtained from \eref{e243} for corresponding
scattering lengths.

It is clear from the figure that it is possible to map 3D results onto 
2D results, but the reverse is not always true.  This is due to the 
different behaviour of the systems for large scattering length, where 
the energy and square radius of the two-body system 
are controlled by the properties of the 
external trap.  
In two dimensions we find that $\langle r^2\rangle=\ell^2$ 
when $a\rightarrow\infty$ and \eref{freq} then tells us that 
we have to choose $\omega=0$, i.e. we have a non-interacting system. 
For three dimensions, the situation is somewhat different as the virial
theorem applied for $a\rightarrow\infty$, tells us that $\langle r^2\rangle=\ell^2/2$.
From \eref{freq} we deduce that $\omega=2\sqrt{2}\omega_0$, i.e.
our harmonic equivalent is an interacting system still.
This also implies that the interaction frequency 
in 3D, $\omega_{}$ in \eref{outputfreq}, can never be smaller than 
$2\sqrt{2}\omega_0$.  In 2D, there is no lower limit for the interaction 
frequency and it eventually vanishes at a large enough scattering 
length where the external field determines all properties.

From \fref{scatlength} we can in principle from a series of 2D
calculations for different scattering lengths extract 3D results.  The
procedure is to start with the desired 3D scattering length, find the
corresponding 2D scattering length, read off the related 2D results for
normal modes, radii, and oscillator energy shift.  The normal modes
and radii are now the 3D results, and the 3D energy is obtained by
finding the related 3D oscillator energy shift from \fref{scatlength}
combined with the expression in \eref{e303}.  The procedure can be reversed provided the desired 2D scattering length is within the range accessible to conversion from 3D (i.e., $a_{2D}/\ell_{2D}\leq 1.14$). 

The hamiltonian we obtain in our oscillator approximation can be decoupled 
by a change of coordinates as discussed above.
 The normal modes can therefore be viewed in one dimension at a time as the
amplitudes with which 
each of the individual particles are moved when the corresponding 
mode is excited on top of the ground state. More explicitly, consider
exciting the $i$'th mode with probability $p$, so that the wavefunction
of the system becomes 
$|\Psi(t)\rangle=\sqrt{1-p^2}|0\rangle+\sqrt{p}e^{-i\bar{\omega}_i t} |i\rangle$,
where $\bar{\omega}_i$ is the frequency of mode $i$. The displacement in
time is then $x_i(t)=\langle\Psi(t)|x|\Psi(t)\rangle=A_{0i}\cos(\bar{\omega}_it)$, 
with amplitude given by $A_{0i}=2\sqrt{p(1-p^2)}\langle 0|x|i\rangle$. We 
illustrate the modes pictorially below.
The fact that the different
spatial directions are decoupled also implies that a spherical
external field produces degenerate modes. This degeneracy can be
lifted by deforming the field but such a symmetry breaking would only
reflect the properties of the field. We therefore consider the
one-dimensional eigenmodes which apply for both $2D$ and $3D$
when using a correspondence of scattering lengths down to $1D$ 
as was done between $2D$ and $3D$ in \fref{scatlength}.

\begin{figure}
\centering
\includegraphics[width=0.7\textwidth]{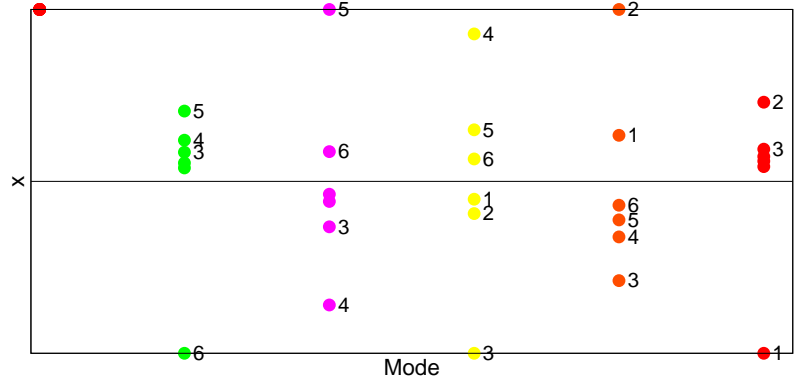}
\caption{Amplitudes of the normal modes in the $x$ direction of a six particle system in 3D calculated with $a/\ell=1$, which is equivalent to a ratio of $0.61$ in the 2D.  The largest amplitude is rescaled to unit magnitude.  Points in the diagram are numbered from one to six with the mass increasing from particles one to six.  Overlapping points are not labeled, and appear in groups usually close to the origin, which is indicate by the black solid horizontal line.   The points oscillate through the equilibrium position (the black solid line) with a frequency of the given normal mode(e.g., $x_i(t)=A_{0i}\cos(\bar\omega_i t)$, where $A_{0i}$ are the amplitudes shown in the figure).  The frequencies from left to right are 1.003, 10.433, 10.447, 10.456, 10.470, and 10.487 in units of $\omega_0$ and are not plotted to scale.  The center of mass mode is the first mode shown.}
\label{3DNM6}
\end{figure}

\begin{figure}
\centering
\includegraphics[width=0.7\textwidth]{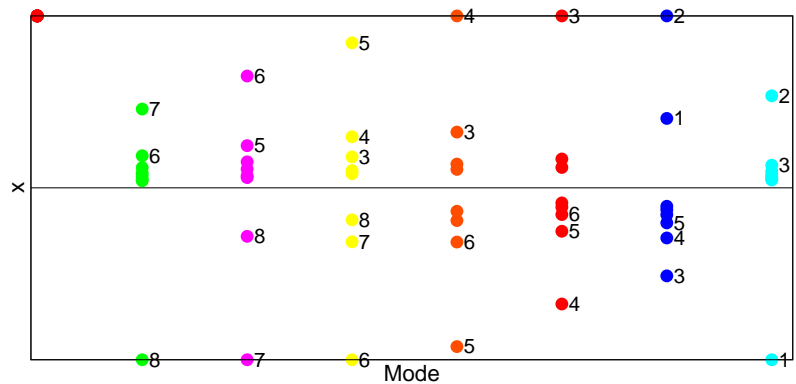}
\caption{Amplitudes of the modes in the $x$ direction of an eight particle system in 3D with $a/\ell=1$, which has an equivalent ratio in 2D of 0.61.  As in \fref{3DNM6}, the largest amplitude is normalized to one unit of length.  Points in the diagram are numbered from one to eight with the mass increasing from particles one to eight (overlapping points are not labeled).  These points oscillate through the equilibrium position, indicated by the solid black horizontal line, with a frequency of the given normal mode.  The frequencies of the normal modes are, from left to right: 1.004, 12.030, 12.040, 12.050, 12.059, 12.069, 12.085, 12.097 in units of $\omega_0$ and are not plotted to scale.  The center of mass mode is the first mode shown.}
\label{3DNM8}
\end{figure}

In  \fref{3DNM6} and \fref{3DNM8} we show a sequence of modes for
six and eight particles respectively.  The dependence on
scattering length is rather weak for both small and large $a/l$.
The general
picture that emerges from breaking the symmetry by mass differences is
first that the energy of the center-of-mass oscillation is maintained.
Second, the energies of the smallest and the largest mode correspond to
oscillation where either the lightest or the heaviest particles move in one
direction while all the others move less and in the opposite direction.  
The remaining
modes with intermediate energies correspond to one particle joining 
the lone particle, moving in the order of increasing mass for each mode, 
though in most cases it appears that only one or two particles are displaced 
significantly, while the remainder stay in a slightly displaced clump.  
It should be noted that the normal modes are highly sensitive to the 
symmetries of the interaction, and how we broke the symmetry in order 
to show non-degenerate normal modes.  In this present case, we
changed the masses slightly to make the particles distinguishable, but still with 
identical interactions between all particles.  We expect that if 
the interactions are changed slightly in a certain 
prescribed manner, then the normal modes will reflect the symmetries 
of that change.

\section{Summary and Conclusion} 

The properties of the quantum mechanical $N$-body system are
determined from the basic one- and two-body interactions. However, in 
general this problem is very hard or impossible to solve.
Here we use an approximate approach which replaces
the interactions with quadratic forms in the coordinates, either by
direct fits of the potentials or by adjusting parameters to reproduce
crucial properties. This approximation allows analytical
investigations of the $N$-body system with the properties expressed in
terms of the two-body characteristics. Having such a direct approach
to the general many-body problem can provide both important analytical
insight and be a valuable benchmark for more intricate methods.

The most general harmonic oscillator potentials to describe one- and
two-body interactions allow analytical solutions of the
$N$-body Schr\"{o}dinger equation.  However, the center-of-mass degree of
freedom requires special attention as we have described above.
We have employed Cartesian coordinates and
one-, two-, and three-dimensional systems are therefore equally simple
to handle.  We have discussed the
properties of the resulting set of decoupled oscillators
and their relation to the initial interactions. More explicitly we calculated
energies, radii, and the one-body density matrix and presented its 
spectrum.

As an application of our formalism, we consider first the simplest case of $N$ bosons in a trap 
interacting through
contact potentials. This was done by using a mapping from the energies and radii
for the two-body system to the oscillator parameters for
the analytical calculations. We note that while our choice of energy and 
radius as the essential parameters to reproduce in the two-body system was
perhaps the most physically reasonable one, other choices are also possible as long
as one has two quantities that fix the oscillator frequency and the shift.
We calculated energies and radii as
function of boson number and the scattering length as a measure of the
interaction strength.  Typically the binding energies increase and the
radii decrease with $N$.  We also discussed characteristic limits for
scattering lengths much smaller and much larger than the length scale
of the external trap. 

An interesting calculation was done for the 
one-body density matrix for which we calculated the dependence of the
lowest eigenvalue with the number of bosons and the scattering length, 
with a careful treatment
removing the center-of-mass degree of freedom. The limits of 
large scattering length gave large condensate fractions, whereas for
small scattering length we found a fragmented state. The former is due
to the near non-interacting system, whereas the latter is caused by the 
strongly bound molecular state which introduces substantial amounts
of correlations in the system. These conclusions apply equally for 
both two and three dimensions.
We also computed the critical number of particles that can form a self bound system of
negative energy.  This number increases with increasing
scattering length. In three dimensions this happens already for a few
particles while in two dimensions more than 50 bosons are needed for
large scattering lengths.  

As a novelty, we considered the normal modes which are characteristic properties of any
system.  However, in the case of the $N$-boson system, there is a large degree of 
degeneracy that can obscure the detailed behaviour.
The ambiguity 
due to degenerate eigenmodes is circumvented by
breaking the symmetry and then approaching the limit of full degeneracy for
the identical boson system.  A different way of breaking the symmetry
is to deform the external field but the resulting eigenmodes then
reflect precisely the chosen deviation from spherical symmetry. We
then computed the one-dimensional oscillatory eigenmodes which can
be related to equivalent values of the scattering lengths in two and three dimensions.
What we found in the normal modes was an interesting tendency for the 
particles to cluster in smaller groups and then perform motion with 
respect to other such groups. This implies that excitations can induce
strong correlations in a many-boson system even when all interactions are
equal. 

The method that we have presented in this report is completely general and 
can be applied to systems in external fields or to self-bound structures. 
The treatment of deformation of the trap or the two-body interaction 
is straightforward and we expect that rotation of the external 
trapping potential poses no problem as well. While we have only treated 
bosons in this work, the extension to fermions is achieved through 
proper (anti-) symmetrization of the wave function and also Bose-Fermi
mixtures are accessible. With the possibility of having displaced 
centers we can also apply the method for split traps, and also 
even more exotic geometries with mixed dimensions which are
under current study within Fermi-Fermi mixtures 
of ultracold atoms \cite{nishida08,nishida09,levinsen09,lampo10}. Our 
method can also be applied to cold polar molecules, in particular 
to the case where two-dimensional confinement is induced to make
layered systems \cite{jin10} where non-trivial two- and three-body bound states 
appear in the bilayer \cite{shih09,arm10,kla10,artem10,wunsch10,wunsch11} that open 
for study of various exotic many-body states in both bilayer and 
multi-layer systems \cite{wang06,wang07,lut09,potter10,pikovski10,zin10}.
The form of the dipolar potential has harmonic oscillator
shape in the inner part when particles are confined in multi-layers
and we therefore expect the the method presented here to readily
provide analytical results valid for intermediate and strong
dipolar strength. The fact that normal modes are easily accessible
with the harmonic method presented makes this even more interesting.
This can help understand the modes of excitation in 
chains and complexes for use in thermodynamic calculations of 
system properties.

In conclusion, the analytic solutions of coupled harmonic oscillators
can be used to study the overall properties of $N$-body systems
and structures can be calculated in general from two-body properties for
different particle number and geometries. This can serve as a valuable
complement to the understand of results obtained with more intricate 
methods or help solve systems that are intractable in other approaches.


\section*{References}
\begin{thebibliography}{99}


\bibitem{goep55} Goeppert-Mayer M and Jensen J H D 1955 \textit{Elementary Theory of Nuclear Shell Structure} (John Wiley \& Sons, Inc., New York)

\bibitem{heyd90} Heyde K L G 1990 \textit{The Nuclear Shell Model}  (Springer, Berlin) 

\bibitem{bloch08} Bloch I, Dalibard J and Zwerger W 2008 \RMP \textbf{80} 885

\bibitem{ludvig2010} Lizana L, Ambj{\"o}rnsson T, Taloni A, Barkai E and Lomholt M A 2010 {\it Phys. Rev. E} \textbf{81} 051118 

\bibitem{zhan08} Zhang J Y, Mitroy J and Varga K 2008 {\it Phys. Rev. A} \textbf{78} 042705

\bibitem{brosens97a} Brosens F, Devreese J T and Lemmens L F 1997  {\it Phys. Rev. E} {\bf 55} 227 
\bibitem{brosens97b} Brosens F, Devreese J T and Lemmens L F 1997  {\it Phys. Rev. E} {\bf 55} 6795 
\bibitem{brosens97e} Brosens F, Devreese J T and Lemmens L F 1997  {\it Phys. Rev. A} {\bf 55} 2453 
\bibitem{brosens97c} Brosens F, Devreese J T and Lemmens L F 1998  {\it Phys. Rev. E} {\bf 57} 3871 
\bibitem{brosens97d} Brosens F, Devreese J T and Lemmens L F 1998  {\it Phys. Rev. E} {\bf 58} 1634 
\bibitem{tempere98a} Tempere J, Brosens F, Lemmens L F and Devreese J T 1998 {\it Phys. Rev. A} {\bf 58} 3180 
\bibitem{lemmens99a} Lemmens L F, Brosens F and Devreese J T 1999  {\it Phys. Rev. A} {\bf 59} 3112
\bibitem{foulon1999} Foulon S, Brosens F, Devreese J T and Lemmens L F 1999 {\it Phys. Rev. E} {\bf 59} 3911
\bibitem{tempere00a} Tempere J, Brosens F, Lemmens L F and Devreese J T 2000 {\it Phys. Rev. A} {\bf 61} 043605 

\bibitem{zalu00} Za\l uska-Kotur M A, Gajda M, Or\l owski A and Mostowski J 2000  {\it Phys. Rev. A} \textbf{61} 033613

\bibitem{yan03} Yan J 2003 {\it J. Stat. Phys.} \textbf{113} 623

\bibitem{gajd06} Gajda M 2006  {\it Phys. Rev. A} \textbf{73} 023603

\bibitem{busch98} Busch T, Englert B G, Rza\.zewski K and Wilkens M 1998 {\it Found. Phys.} \textbf{28} 548

\bibitem{stoferle06} St{\"o}ferle T, Moritz H, G{\"u}nter K, K{\"o}hl M and Esslinger T 2006 \PRL {\bf 96} 030401

\bibitem{yang60} Yang C N 1962 \RMP {\bf 34} 694

\bibitem{zinner08} Zinner N T and Jensen A S 2008 {\it Phys. Rev. C} {\bf 78} 041306(R)

\bibitem{pit00}  Pethick C J and L. P. Pitaevskii L P 2000 {\it Phys. Rev. A} {\bf 62} 033609

\bibitem{gajd00} Gajda M, Za{\l}uska-Kotur M A and Mostowski J 2000 \jpb \textbf{33} 4003

\bibitem{gattobigio2009} Gattobigio M, Kievsky A, Viviani M and Barletta P. 2009 {\it Phys. Rev. A} {\bf 79} 032513

\bibitem{stet07} Stetcu I, Barrett B R, van Kolck U and Vary J P 2007 {\it Phys. Rev. A} \textbf{76} 063613

\bibitem{alhassid08} Alhassid Y, Bertsch G F and Fang L 2008 \PRL {\bf 100} 230401

\bibitem{rotu10} Rotureau J, Stetcu I, Barrett B R, Birse M C and van Kolck U 2010 {\it Phys. Rev. A} \textbf{82} 032711

\bibitem{stet10} Stetcu I, Rotureau J, Barrett B R and van Kolck U 2010 {\it Ann. Phys.} \textbf{325} 1644

\bibitem{zinner09} Zinner N T, M{\o}lmer K, {\"O}zen C, Dean D J and Langanke K 2009 {\it Phys. Rev. A} {\bf 80} 013613

\bibitem{cem09} {\"O}zen C and Zinner N T 2009 {\it Preprint} arXiv:0902.4725v1 

\bibitem{carlson03} Carlson J, Chang S Y, Pandharipande V R and Schmidt K E 2003 \PRL {\bf 91} 050401

\bibitem{chang07} Chang S Y and Bertsch G F 2007 {\it Phys. Rev. A} {\bf 76} 021603(R)

\bibitem{gezerlis08} Gezerlis A and Carlson J 2008 {\it Phys. Rev. C} {\bf 77} 032801(R)

\bibitem{ole02} S{\o}rensen O, Fedorov D V and Jensen A S 2002 \PRL {\bf 89} 173002 

\bibitem{ole03}  S{\o}rensen O, Fedorov D V and Jensen A S 2003 {\it Phys. Rev. A} {\bf 68} 063618

\bibitem{ole04a} S{\o}rensen O, Fedorov D V and Jensen A S 2004 {\it Phys. Rev. A} {\bf 70} 013610

\bibitem{han06} Hanna G J and Blume D 2006 {\it Phys. Rev. A} {\bf 74} 063604

\bibitem{thoger07} Th{\o}gersen M, Fedorov D V and Jensen A S 2007 {\it Europhysics Letters} {\bf 79} 40002

\bibitem{lim80} Lim T K, Nakaichi S, Akaishi Y and Tanaka H 1980 {\it Phys. Rev. A} {\bf 22} 28

\bibitem{ole04b} Guangze H, S{\o}rensen O, Jensen A S and Fedorov D V 2004 {\it Phys. Rev. A} {\bf 70} 013609

\bibitem{blume05} Blume D 2005 {\it Phys. Rev. B} {\bf 72} 094510

\bibitem{chri09} Christensson J, Forss{\'e}n C, {\AA}berg S and Reimann S M 2009 {\it Phys. Rev. A} \textbf{79} 012707

\bibitem{ront09} Rontani M, Armstrong J R, Yu Y, {\AA}berg S and Reimann S M 2009 \PRL \textbf{102} 060401

\bibitem{liu10} Liu X J, Hu H and Drummond P D 2010 {\it Phys. Rev. B} \textbf{82} 054524

\bibitem{ront08} Rontani M, {\AA}berg S and Reimann S M 2008 {\it Preprint} arXiv:0810.4305v1

\bibitem{bert06} Bertelsen J F and M{\o}lmer K 2006 {\it Phys. Rev. A} \textbf{73} 013811

\bibitem{bert07} Bertelsen J F and M{\o}lmer K 2007 {\it Phys. Rev. A} \textbf{76} 043615

\bibitem{lovelace87} Lovelace R V E and Tommila T J 1987 {\it Phys. Rev. A} {\bf 35} 3597

\bibitem{baym96} Baym G and Pethick C J 1996 \PRL {\bf 76} 6

\bibitem{fu03} Fu H, Wang Y and Gao B 2003 {\it Phys. Rev. A} {\bf 67} 053612

\bibitem{zinner09b} Zinner N T and Th{\o}gersen M 2009 {\it Phys. Rev. A} {\bf 80}, 023607

\bibitem{thoger09} Th{\o}gersen M, Zinner N T and Jensen A S 2009 {\it Phys. Rev. A} {\bf 80} 043625

\bibitem{bagnato91} Bagnato V and Kleppner D 1991 {\it Phys. Rev. A} {\bf 44} 7439

\bibitem{nishida08} Nishida Y and Tan S 2008 \PRL {\bf 101} 170401 

\bibitem{nishida09} Nishida Y and Tan S 2009 {\it Phys. Rev. A} {\bf 79} 060701(R)

\bibitem{levinsen09} Levinsen J, Tiecke T G, Walraven J T M and Petrov D S 2009 \PRL {\bf 103} 153202

\bibitem{lampo10} Lamporesi G {\it et al.} 2010 \PRL {\bf 104} 153202

\bibitem{jin10} de Miranda M H G {\it et al.} 2011 {\it Nature Phys.} {\bf 7} 502  

\bibitem{shih09} Shih S M and Wang D W 2009 {\it Phys. Rev. A} {\bf 79} 065603

\bibitem{arm10} Armstrong J R, Zinner N T, Fedorov D V and Jensen A S 2010 {\it Europhysics Letters} {\bf 91} 16001

\bibitem{kla10} Klawunn M, Pivovski A and Santos L 2010 {\it Phys. Rev. A} {\bf 82} 044701

\bibitem{artem10} Volosniev A, Zinner N T, Fedorov D V, Jensen A S and Wunsch B 2011 {\it J. Phys. B} {\bf 44} 125301 

\bibitem{wunsch10} Wunsch B {\it et al.} 2011 \PRL {\bf 107} 073201

\bibitem{wunsch11} Zinner N T {\it et al.} 2011 {\it Phys. Rev. A} {\bf 84} 063606

\bibitem{wang06} Wang D W, Lukin M D and Demler E 2006 \PRL {\bf 97} 180413

\bibitem{wang07} Wang D W 2007 \PRL {\bf 98} 060403

\bibitem{lut09} Lutchyn R M, Rossi E and Das Sarma S 2010 {\it Phys. Rev. A} {\bf 82} 061604(R) 

\bibitem{potter10} Potter A C {\it et al.} 2010 \PRL {\bf 105} 220406

\bibitem{pikovski10} Pikovski A, Klawunn M, Shlyapnikov G V and Santos L  2010 \PRL {\bf 105} 215302

\bibitem{zin10} Zinner N T, Wunsch B, Pekker D and Wang D W 2012 {\it Phys. Rev. A} {\bf 85} 013603




\end{thebibliography}
\end{document}